# High-dimensional time-frequency entanglement in a singly-filtered biphoton frequency comb

Xiang Cheng[1†*], Kai-Chi Chang[1†*], Murat Can Sarihan[1], Andrew Mueller[2,3], Maria Spiropulu[4], Matthew D. Shaw[2], Boris Korzh[2], Andrei Faraon[5], Franco N. C. Wong[6], Jeffrey H. Shapiro[6], and Chee Wei Wong[1*]

[1]Fang Lu Mesoscopic Optics and Quantum Electronics Laboratory, Department of Electrical and Computer Engineering, University of California, Los Angeles, CA 90095, USA

[2]Jet Propulsion Laboratory, California Institute of Technology, 4800 Oak Grove Dr., Pasadena, CA 91109, USA

[3]Applied Physics, California Institute of Technology, 1200 E California Blvd, Pasadena, CA 91125, USA

[4]Division of Physics, Mathematics and Astronomy, California Institute of Technology, Pasadena, CA 91125, USA

[5]Thomas J. Watson, Sr., Laboratory of Applied Physics, California Institute of Technology, Pasadena, CA 91125, USA

[6]Research Laboratory of Electronics, Massachusetts Institute of Technology, Cambridge, MA 02139, USA

†These authors contributed equally to this work

*email: chengxiang@ucla.edu; uclakcchang@ucla.edu; cheewei.wong@ucla.edu

**High-dimensional quantum entanglement is a cornerstone for advanced technology that could lead to practical large-scale quantum systems with higher data capacity and noise tolerance [1, 2], to fault-tolerant quantum computing [3-6], and to distributed quantum networks [7-9]. The recently developed biphoton frequency comb (BFC) [10-14], with its capacity for large-scale information encoding in its spectral and temporal quantum modes, provides a powerful platform for high-dimensional quantum communication [15] and information processing [16, 17]. Here we first generate a high-dimensional BFC via spontaneous parametric down-conversion in a singly-filtered configuration by spectrally shaping only the signal photons with a Fabry-Pérot cavity, imbuing their idler-photon companions with the same comb signature due to conservation of energy in the biphoton generation process. The coherence of the BFC state is examined by low timing-jitter**

detectors to resolve the temporal correlation. We witness the high-dimensional energy-time entanglement through Franson-interference recurrences, for the first time in a singly-filtered BFC, with 99.46 ± 1.09% visibility over integer time bins. Energy-time entanglement between different frequency-bin pairs of the singly-filtered BFC is subsequently verified by Franson interferometry with an average visibility up to 98.03 ± 1.10% for 5 symmetric frequency-bin pairs. Second, frequency- and temporal- entanglement of the singly-filtered BFC is then characterized by binned Schmidt-mode decompositions of the joint spectral and temporal intensities, with respectively a measured Schmidt number of 4.17 across 5 frequency bins and a measured Schmidt number of 13.11 inferred from Franson-interference recurrences across 16 time bins. Thirdly, we distribute the entanglement of the high-dimensional singly-filtered BFC state asymmetrically over a 10 km fiber link, with a long-distance visibility up to 98.81 ± 0.61% for the time-bins. Post-distribution, we obtain a Schmidt number of 12.99, lower bounding the time-bin dimensions of our distributed singly-filtered BFC state to be at least 168. An averaged Franson visibilities up to 96.70 ± 1.93% is obtained across 5 frequency bins after 10 km distribution, which verifies the high-quality entanglement distribution of our high-dimensional time-frequency entangled BFC state. Our demonstrations of high-dimensional entanglement and entanglement distribution show that the singly-filtered quantum frequency comb provides a versatile platform for high-efficiency quantum information processing and high-capacity quantum networks, also serving as a valuable tool for fundamental research on locality violations in high-dimensional systems.

High-dimensional quantum entanglement is a cornerstone for advanced technology enabling large-scale noise-tolerant quantum systems, fault-tolerant quantum computing, and distributed quantum networks. The recently developed biphoton frequency comb (BFC) provides a powerful platform for high-dimensional quantum information processing in its spectral and temporal quantum modes. Here we propose and generate a singly-filtered high-dimensional BFC via spontaneous parametric down-conversion by spectrally shaping only the signal photons with a Fabry-Pérot cavity. High-dimensional energy-time entanglement is verified through Franson-interference recurrences and temporal correlation with low-jitter detectors. Frequency- and temporal- entanglement of our singly-filtered BFC is then quantified by Schmidt mode decomposition. Subsequently, we distribute the high-dimensional singly-filtered BFC state over a 10 km fiber link with a post-distribution time-bin dimension lower bounded to be at least 168. Our demonstrations of high-dimensional entanglement and entanglement distribution show the singly-filtered quantum frequency comb's capability for high-efficiency quantum information processing and high-capacity quantum networks.

**Introduction**

Quantum entanglement, referred to as "spooky action at a distance" [18], has enabled tremendous advances in both fundamental science and engineering technologies [19, 20]. As a technological resource, quantum entanglement has revolutionized communications [21, 22], computation [23, 24], metrology [25], and sensing [26]. High-dimensional entanglement, owing to its higher information capacity and resilience to noise, has been proposed for noise-resilient large-alphabet quantum key distribution [27] and universal photonic quantum computation [28]. High-dimensional quantum states of entangled photons have already been demonstrated in a variety of degrees-of-freedom, such as spatial modes [29], orbital angular momentum [30], optical frequency [31] and time bins [32]. Recently demonstrated BFCs, which can carry information in their spectro-temporal quantum modes, are ideal candidates for high-dimensional quantum systems. Such BFCs can be produced by spontaneous parametric down-conversion (SPDC) with post-generation filtering [10, 13, 14, 17], optical parametric oscillation (OPOs) operating far below threshold [33, 34], or spontaneous four-wave mixing in integrated microring resonators [11, 31, 35]. The first approach is realized by sending SPDC-generated biphotons through a cavity, while the second relies on cavity-enhanced SPDC, in which the biphoton generation is both enhanced and spectrally confined by the cavity modes. The third approach utilizes third-order nonlinearity to generate photon entanglement over a broad range of frequency modes in a chip-scalable platform. These BFCs approaches are usually doubly resonant, viz., the signal and idler's spectra are simultaneously tightly confined to the cavity modes. On the other hand, the singly-resonant BFC has been studied, theoretically, for a sub-threshold OPO in which only the signal photons are resonated by the OPO cavity [36]. Owing to the entanglement between signal and idler photons from biphoton generation process, the idler photons will then exhibit the comb-like spectrum corresponding to the OPO cavity's internal mode structure [37]. Recently, such a singly-resonant BFC has been demonstrated to generate frequency-multiplexed photon pairs over 1,000 frequency modes [38], and also highly multi-mode polarization-entangled photon pairs by means of a Sagnac interferometer [39]. However, a typical OPO configuration employs a cavity that is a few cm long, to accommodate the nonlinear crystal used for SPDC. Hence its BFC's free spectral range (FSR) is limited to a few GHz. That bandwidth is not compatible with off-the-shelf dense wavelength-division multiplexer/demultiplexer devices, whose channel bandwidths are 50 or 100 GHz. Although the singly-resonant BFC offers a promising platform for quantum information processing, its high-dimensional time-frequency entanglement has not been carefully investigated. Moreover, distributing its high-dimensional entanglement [31, 40, 41], which is a critical precursor to its enabling large-scale high-dimensional quantum communication and distributed networks, remains a challenge. Indeed, high-dimensional time-frequency entanglement distribution has yet to be demonstrated for the singly-resonant BFC.

Here we propose a flexible approach to generate a singly-filtered BFC state with the same temporal and spectral properties as the OPO based singly-resonant BFC by spectrally shaping only the signal photon of the SPDC-generated pair with a Fabry-Pérot cavity. We observe that this singly-filtered BFC exhibits the same temporal correlation as the singly-resonant BFC [36, 38, 39]. Moreover, high-dimensional energy-time entanglement of such singly-filtered BFC is verified for the first time, via Franson-interference recurrences over 16 times-bins. Specially, we verify the spectral phase coherence by resolving the periodic oscillations in the cross-correlation with single-sided decay using state-of-the-art low timing jitter superconducting nanowire single-photon detectors (SNSPDs), supporting the high-dimensional frequency-bin entanglement of our BFC state. The joint spectral intensity of the singly-filtered BFC is measured and analyzed through Schmidt mode decomposition with 4.17 Schmidt number over 5 frequency bins. Recurrences of the Franson-interference further enable the mapping of the BFC's time-binned joint temporal intensity, shown to have a Schmidt number of 13.11 over 16 time bins, concurring with our high-dimensional time-bin entangled state. Furthermore, entanglement distribution of the singly-filtered BFC's high-dimensional state is demonstrated over a 10 km optical fiber link in an asymmetric configuration. The post-distribution BFC state's high-dimensional time-frequency entanglement is examined via non-local interferometry, with up to $98.81 \pm 0.61\%$ visibility achieved recurrently over 16 time-bins and with an averaged $96.70 \pm 1.93\%$ visibility across 5 frequency-bin pairs. We also estimate the Schmidt number to be 12.99 from the distributed non-local quantum interference revivals, which lower bounds the time-binned Hilbert-space dimensionality to be at least 168. Furthermore, we demonstrate proof-of-principle high-dimensional quantum key distribution with our singly-filtered BFC, exploiting the lower filtering loss of the singly-filtered configuration. This first high-dimensional time-frequency entanglement distribution paves the pathway in constructing practical long-distance quantum networks.

**Results**

**Generation and spectro-temporal characterization of a singly-filtered BFC**

Figure 1a illustrates the experimental setup to generate and characterize the singly-filtered BFC. The entangled photon pairs are generated by a 16-mm long type-II periodically-poled KTiOPO$_4$ (ppKTP) waveguide (AdvR Inc.) that was integrated in a fiber package for high fluence and efficiency [10, 13]. A 658 nm Fabry-Pérot laser diode, stabilized by self-injection locking through double-pass first-order diffraction feedback using an external grating, is used to pump the ppKTP waveguide. The generated biphotons are orthogonally polarized and frequency degenerate at 1316 nm with ≈ 245 GHz full-width half-maximum (FWHM) bandwidth. The residual pump photons are removed by a long-pass filter (LPF). A 1.3 nm bandpass filter (BPF), i.e., 225 GHz FWHM bandwidth, is used to further clean the biphoton spectrum. Then the signal and idler photons are separated efficiently by a polarizing beam splitter (PBS) due to the type-II phase matching. The singly-filtered BFC is generated by passing only the signal photons through a fiber Fabry-Pérot cavity (FFPC) (Luna Inc.). The idler photons, although not confined in a cavity, still exhibit a comb-like spectrum when heralded by signal-photon detections, due to frequency entanglement [36]. The FFPC, which has a 45.32 GHz FSR and 1.56 GHz FWHM bandwidth, is stabilized with a high-performance temperature controller. We note that if the FFPC is used in a doubly-filtered configuration – when both the signal and idler photons are filtered – the FFPC's polarization birefringence would result in different post-filtering spectra for the signal and idler photons. However, in the present singly-filtered configuration, only the signal photons pass through the FFPC and we further solve the requirement for polarization birefringence elimination.

We first characterize the temporal signature of the singly-filtered BFC. Our singly-filtered configuration generates the (unnormalized) BFC state whose frequency-domain representation can be usefully approximated as:

$$|\psi\rangle = \sum_{m=-N}^{N} \int d\Omega\, f(\Omega - m\Delta\Omega)\mathrm{sinc}(A\Omega)\, \hat{a}_H^\dagger(\frac{\omega_p}{2} + \Omega)\hat{a}_V^\dagger(\frac{\omega_p}{2} - \Omega)|0\rangle; \quad (1)$$

Here: $\hat{a}_H^\dagger$ and $\hat{a}_V^\dagger$ are the creation operators for horizontally (signal) and vertically (idler) polarized photons; $\omega_p$ is the pump frequency; the sinc function is the SPDC's phase-matching function with $A = 1.39/\pi B_{PM}$ for $B_{PM} = 245$ GHz being the FWHM phase-matching bandwidth; $\Delta\Omega$ is the FFPC's FSR in rad s$^{-1}$; $\Omega$ is the detuning of the SPDC biphotons from their center frequency; $2N + 1 = 5$ is the number of the cavity lines passed by the bandwidth-limiting filter; and $f(\Omega - m\Delta\Omega)$ is the spectral amplitude of the FFPC's $m$th cavity resonance, with $f(\Omega) = 1/[\Delta\omega + i\Omega]$, i.e., a Lorentzian transmission whose FWHM linewidth is $2\Delta\omega$. The temporal representation of the singly-filtered BFC that is the dual of Eq (1) is then:

$$|\psi\rangle = \int_0^\infty d\tau \exp(-\Delta\omega\tau) \sum_{m=-N}^{N} \text{sinc}(Am\Delta\Omega)\cos(m\Delta\Omega\tau)\hat{a}_H^\dagger(t+\tau)\hat{a}_V^\dagger(t)|0\rangle, \qquad (2)$$

where we have used $\Delta\Omega/2\pi \ll B_{PM}$. The state's temporal behavior then has recurrences with repetition period $\Delta T = 2\pi/\Delta\Omega \approx 22.1$ ps, i.e., the cavity's round-trip time. We note that for the doubly-filtered BFC, the integral in eq. (2) spans from negative to positive, which results in the symmetric temporal behaviors [10, 13].

Figure 1b shows the normalized second-order cross-correlation function as a function of relative delay between signal and idler photons (defined as $\tau = t_{idler} - t_{signal}$), measured with superconducting nanowire single-photon detectors (SNSPDs, ≈ 80% detection efficiency, PhotonSpot Inc.). A single-sided exponential decay of the cross-correlation function is the temporal signature of a singly-resonant BFC [36-38]. The measured second-order cross-correlation function matches well with our theoretical calculations assuming the detectors' combined root-mean-square timing jitter is $t_j = 3.4\Delta T = 74$ ps. Here the temporal oscillation signature of the cross-correlation function is not fully resolved because the timing jitter is more than 3 times the repetition time period. Low time-jitter SNSPDs have been recently demonstrated, which enable better temporal resolution [42]. Figure 1b shows that with $t_j = 0.35\Delta T = 7.8$ ps the temporal oscillation profile of the 45.32 GHz singly-filtered BFC can be observed; with $t_j = 0.035\Delta T = 0.78$ ps, that signature is fully resolved, showing an oscillation period of 22.1 ps.

Next we characterize the frequency correlation of our 45.32 GHz singly-filtered BFC through joint spectral intensity (JSI) measurements. We use a 2 mW pump for SPDC generation to minimize multiphoton emission and reduce cross-talk between frequency bins. The JSI of our singly-filtered BFC is measured using a pair of tunable BPFs for signal and idler photons respectively. The BPFs have 300 pm bandwidth, which are able to select only one frequency bin of our BFC. Signal-idler coincidence counts are recorded while the tunable BPFs are set to different combinations of the signal-idler frequency-bin pairs. Within the 245 GHz SPDC bandwidth, 5 frequency bins can be examined for the signal and idler of our 45.32 GHz singly-filtered BFC. We sweep the BPFs from -2 to +2 frequency bins, with the $0^{th}$ frequency bin indicating the SPDC's center frequency, i.e., half the pump frequency. High values of photon coincidences are measured only for symmetric frequency-bin pairs, shown as the diagonal elements of the frequency correlation matrix in Figure 1c. This behavior reveals the frequency correlation of the singly-filtered BFC, and is a characteristic of frequency-bin entanglement. We note that the coincidence counts fall off at frequency bin pairs away from the central bin, which results from the sinc-squared spectra of the SPDC biphotons prior to signal filtering.

For each symmetric frequency-bin pair, we then measure the signal-heralded second-order auto-correlation $g^{(2)}(0)$ of the idler photons. The measurement is performed using a Hanbury-Brown and Twiss (HBT) interferometer in which the idler light from the singly-filtered BFC is divided into two paths by a 50:50 beam splitter for auto-correlation measurement heralded by signal-photon detections. A pair of tunable BPFs (300 pm bandwidth) is placed before the HBT interferometer to select different frequency bins. The second-order auto-correlation $g^{(2)}(0)$ is measured by recording the three-fold coincidence counts between the HBT interferometer's output ports and the signal photons within a 2 ns duration coincidence window. Detecting a signal photon heralds the appearance of the idler photon, which exhibits non-classical anti-bunching behavior. Figure 1d shows the signal-heralded $g^{(2)}(0)$ versus pump power for five symmetric frequency-bin pairs. The pump power is set to 1.3 mW, 1.8 mW and 2.4 mW respectively. At low pump power, the heralded $g^{(2)}(0)$ values for all five frequency-bin pairs are below 0.1, showing high single-photon purity of the frequency-filtered states from our singly-filtered BFC. We observe that the heralded $g^{(2)}(0)$ is proportional to the pump power, due to the Poisson statistics of the SPDC emission. With increased pump power, we note that the heralded $g^{(2)}(0)$ for each frequency-bin pair increases because of multi-pair emissions. At higher pump power, the heralded $g^{(2)}(0)$ for $S_0$&$I_0$ ($S_0$ and $I_0$ denote the central frequency bin for signal and idler photons, respectively) frequency-bin pair is still below 0.1. The heralded $g^{(2)}(0)$ values for $S_{+2}$& $I_{-2}$ and $S_{-2}$&$I_{+2}$ increase faster than is the case for $S_{+1}$&$I_{-1}$ and $S_{-1}$&$I_{+1}$ when the pump power is increased. This may be due to the coincidence counts' fall off in Figure 1c, which implies that the frequency bin pairs away from the degeneracy will have worse signal-to-noise-ratio with increasing noise photons introduced by the stronger pump. In addition, we measured the heralded $g^{(2)}(0)$ for the singly-filtered BFC state without selection of a frequency-bin pair by using only a broadband BPF (225 GHz bandwidth) to clean up the biphoton spectrum. A signal-heralded $g^{(2)}(0) \approx 0.154$ is measured for our singly- filtered BFC with 0.2 mW pump power. We also obtain a $g^{(2)}(0) \approx 0.130$ with 0.6 mW pump power by sending the signal photons to the HBT interferometer while all the idler photons are sent to an SNSPD whose detections provided heralding for the three-fold coincidence measurement. Both heralded $g^{(2)}(0)$ values are well below the classical threshold, demonstrating a high-purity heralded single-photon state preparation from our singly- filtered BFC.

**High-dimensional energy-time entanglement witnessed via intrinsic temporal oscillations and Franson-interference recurrences**

We verify the coherence of our high-dimensional singly-filtered BFC states via temporal second-order cross-correlation. We extract the two-photon time-correlation from joint temporal intensity (JTI), namely, the temporal cross-correlation measurements. In order to resolve the temporal correlation of our singly- filtered BFC, the effective timing jitter should be equal to or less than the cavity round-trip time. We employ two state-of-the-art impedance-matched differential SNSPDs with low timing jitter [43] while preserving a moderate detection efficiency, in combination with a multi-channel low-jitter time tagger (Swabian Tagger X) to perform the cross-correlation measurements between the signal and idler photons. First we characterize the combined system timing jitter of the two differential SNSPDs using our SPDC photon source – the jitter adds in quadrature – and obtain a full-width at half maximum of $\approx 21.6$ ps from the cross-correlation function, which is comparable to our 45.32 GHz cavity round-trip time. This allows us to measure the temporal correlation oscillations of our singly-filtered BFC and observe distinct correlation peaks, as shown in the datapoints of Figure 2a. This periodic oscillation arises from the coherent interference of biphotons' different frequency modes, with a temporal spacing of 22.0 ps (corresponds to 45.45 GHz FSR), matching well with the cavity round-trip time of 22.1 ps. Figure 2b shows another cross-correlation measurement with a 15.15 GHz cavity. The temporal spacing in the correlation peaks is 66.8 ps, implying a 14.97 GHz cavity FSR and matching our selected cavity. Supported by our exact theory on the temporal oscillations with the phase-sensitive cross-spectrum of the post-filtered baseband field operator in the presence of detector jitter, the measured intrinsic temporal oscillations demonstrate that our singly-filtered BFC is coherently generated in high-dimensions and with a flat spectral phase [39, 44].

We next characterize the high-dimensional energy-time entanglement of a singly-filtered BFC for the first time by means of its Franson-interference recurrences. We use a 1.3 nm bandwidth BPF to clean the SPDC spectrum before separating the signal and idler photons with a PBS. The signal photons are passed through a 45.32 GHz FFPC to generate our singly-filtered BFC. The signal and idler photons are sent to a fiber-based Franson interferometer comprised of two unbalanced Michelson interferometers, as shown in Figure 1a, that are temperature-controlled by Peltier modules and enclosed in a multilayered thermally-insulated housing with active temperature stabilization for long-term phase stability. The tunable delay line in arm 1 selects the different time-bins, while the thermal heater in arm 2 fine-tunes the relative phase delay between two arms to obtain the interference fringes. The coincidence counts are recorded in a 2 ns duration time window with accidental coincidences subtracted. Figure 2c shows the first observed Franson-interference recurrences from a singly-filtered BFC. Counterintuitively, with its cross-correlation function being single-sided, the Franson-interference recurrences still span both positive and negative relative delays. This originates from the overlap integral between the singly-filtered BFC's phase-sensitive cross-correlation and the it's delay-shifted counterpart. These recurrences are only observed at periodically recurring time bins, whereas no interference is observed between time bins. The measured period of the interference recurrences is 22.1 ps, which corresponds exactly to the round-trip time of the 45.32 GHz cavity ($\Delta T = 2\pi/\Delta\Omega$). The Franson interference fringes are measured by selecting the relative (arm 1 minus arm 2) delay between the signal and idler from 0 (0[th] time-bin) to 15 (15[th] time-bin) cavity round-trip times, limited by the 360 ps tuning range of the delay line in arm 1. We obtain a high Franson interference visibility of 99.46% at the 0[th] time-bin, which is sufficiently high to violate the Clauser-Horne-Shimony-Holt Bell inequality [45].

The Franson interference visibilities decrease as the relative delay increases, i.e., at higher order time-bins, which results from the SPDC's sinc-function phase matching and the FFPC's (approximately) Lorentzian lineshape. The measured Franson-interference recurrences agree well with our theoretical prediction, as shown in Figure 2d. We note that, unlike the symmetric Franson interference fringe envelope for the doubly-filtered BFC [10, 13], our model for the singly-filtered BFC exhibits asymmetric behavior. The Franson visibilities at positive time-bins decrease faster than those at negative time-bins, originating from the *asymmetric* temporal profile of the cross-correlation function. Only the signal photon is resonant with the cavity while the idler photon does not experience the cavity mode (i.e. singly-filtered). Thus the signal photon can only exit the FFPC after $(n + 1/2)$ round trips, where $n$ is a non-negative integer. Therefore, despite being created in time coincidence with its idler companion, the signal photon arrives after an additional time delay with respect to the idler photon, leading to the cross-correlation asymmetry. The measured Franson visibilities are in good agreement with our theoretical calculation, as shown in Figure 2e. We compare the Franson visibility fitting of our singly-filtered BFC with that for doubly-filtered BFC [13] and observe faster decay for the singly-filtered case. For doubly-filtered BFC, whose signal and idler photons are both filtered by the FFPC, its frequency-domain biphoton wave function decays faster than the singly-filtered BFC; while for the singly-filtered case, only the signal photons are filtered, which consequently, by Fourier duality, results in a faster decay of its time-domain biphoton wave functions than the doubly-filtered case.

Based on the measured Franson-interference recurrences, we quantify the time-bin entanglement of our singly resonant BFC through Schmidt mode decomposition. We extracted the Schmidt eigenvalues for each time-bin by a parametric fitting of the experimental data as shown in Figure 2e. By summing up the eigenvalues from each time bin, a Schmidt-number lower bound of 13.11 was obtained. Compared to the doubly-filtered case with a reported time-bin Schmidt number of 18.30 [13], the Schmidt number for time-binned singly-filtered BFC state is lower as a result of its Franson-interference recurrences having faster visibility decay.

**Characterizing energy-time entanglement between frequency-bin pairs via Franson interference**

Here we characterize the energy-time entanglement between different frequency-bin pairs of the singly-filtered BFC. We use a pair of narrowband tunable BPFs with 300 pm bandwidth to select signal and idler frequency bins of our 45.32 GHz singly-filtered BFC. The filtered photons are then sent to the two arms of the Franson interferometer, as shown in Figure 1a. We sweep the two tunable BPFs to select frequency bins from -2 to 2 according to the previous JSI measurement in Figure 1c, collecting the Franson interference fringes by fine tuning the relative delay between the two arms using the thermal heater in arm 2. The tunable delay line in arm 1 is fixed at the $0^{th}$ time-bin delay to obtain optimum Franson interference, with the optimized constructive and destructive Franson interference. Figure 3a shows the Franson interference between symmetric frequency-bin pairs, which are the diagonal elements of the frequency correlation matrix (Figure 1c). A high Franson interference visibility of 99.66 ± 1.67%, after subtracting accidental coincidences, is observed for the central frequency-bin pair $S_0$&$I_0$, a result that agrees well with our singly-filtered BFC being a high-purity biphoton state based on its heralded $g^{(2)}(0)$ measurement in Figure 1d. We also obtain high interference visibilities for other symmetric frequency-bin pairs, with an averaged Franson visibility of 98.03 ± 1.10% for the 5 frequency-bin pairs. We note that the visibility of the frequency-binned Franson interference increases slightly compared to the temporal Franson-interference recurrences because of the narrowband filtering of the singly-filtered BFC.

In addition to the symmetric frequency-bin pairs, we measure the Franson interference for the asymmetric frequency-bin pairs, i.e., the off-diagonal elements of the JSI matrix. We map the measured Franson interference visibilities for all frequency-bin pairs within the SPDC bandwidth in Figure 3b. We observe that only the diagonal elements in Figure 3b show high visibilities, which correspond to symmetric frequency-bin pairs, while the off-diagonal elements, representing asymmetric frequency-bin pairs, exhibit no significant interference. This behavior demonstrates that the energy-time entanglement will only occur between correlated spectral modes, which exhibit time-frequency correlations due to energy conservation from SPDC. Moreover, when we adjust the relative delay away from the central time-bin by sweeping the tunable delay line in Franson interferometer's arm 1, there is no interference recurrence in the other time bins. This absence of interference occurs because the narrowband tunable BPFs strictly limit the signal and idler photons to narrow single-peak passbands. Thus, with no beating between multiple spectral peaks, the signal and idler's temporal cross correlation has an exponential decay without oscillation [38, 39]. Figure 3c's inset is a zoom-in of the central cross-correlation peak of the constructive Franson interference. We observe a single-sided exponential time decay of the cross-correlation function, akin to Figure 1b, illustrating the temporal signature of a singly-filtered BFC.

We note that JSI does not contain any phase information required to prove high-dimensional frequency-bin entanglement [46]. Under the assumption of near-pure-state SPDC biphoton generation based on standard perturbation theory [13], as verified for pulse-pumped SPDC by Kuzucu et al.'s JSI/JTI measurements [47], the filtered SPDC source should emit nearly nonseparable-state BFC biphotons. Indeed, our temporal cross-correlation measurement resolves the temporal structure of the high-dimensional BFC, providing JTI information that verifies a phase coherent BFC state generation [44]. Schmidt mode decomposition based on the JSI measurements is then examined to quantify the frequency-bin entanglement of our 45.32 GHz singly-filtered BFC. By extracting the Schmidt eigenvalues from the measured frequency correlation matrix, we obtain the Schmidt number for the frequency-bin entangled state of our singly-filtered BFC as shown in Figure 3d. This parameter describes the lowest number of Schmidt modes in a bipartite system, and therefore gives a lower bound on its effective dimensionality. We obtain the Schmidt number by summing individual Schmidt eigenvalues for each frequency bin, measured to be 4.17 for the five symmetric frequency-bin pairs. Thus, we lower bound the Hilbert-space dimensionality of our frequency-binned singly resonant BFC state to be at least 16 (= 4.17 × 4.17). Higher Schmidt number of a BFC has been demonstrated in integrated platform, with a reported Schmidt number of 20 for frequency modes [48]. The Schmidt number of our singly-filtered BFC can also be increased by using a broader band SPDC source or cavity with smaller FSR. Note that the diagonal elements of the frequency correlation matrix exhibit a decreasing-envelope behavior of our singly-filtered BFC, which leads to the imperfect Schmidt number compared to the ideal case. This behavior can be avoided by utilizing a biphoton source with a flat-top SPDC spectrum and an FFPC with flat-top transmission. In that case, the generated BFC will have equal amplitude for each frequency bin, with resultant Schmidt eigenvalues for symmetric frequency-bin pairs to be consequently equal. This arrangement would lead to a maximum Schmidt number, equaling the number of frequency modes of the frequency-entangled pure state [49]. The Franson-interference visibilities for five symmetric frequency-bin pairs are also plotted together with the Schmidt eigenvalues in Figure 3d. We note that the Franson interference visibilities change in accordance with the eigenvalues at each frequency-bin pair, which agrees well with the trend of the heralded $g^{(2)}(0)$ for each frequency-bin pair.

**High-dimensional time-frequency entanglement distribution of a singly-filtered BFC**

Extending from the spectro-temporal entanglement of our singly-filtered BFC and the high-dimensional time-frequency entanglement verification, we asymmetrically distributed the high-dimensional BFC states through a 10 km fiber link to demonstrate our singly-filtered BFC is suitable for long distance quantum communication. The experimental scheme for high-dimensional entanglement distribution is shown in Figure 4a. The singly-filtered BFC is generated by placing the FFPC on the signal channel after the PBS. The signal photons are then sent to the Franson interferometer's arm 2, a local unbalanced Michelson interferometer arranged to act as a Mach-Zehnder interferometer, while the idler photons propagate through 10 km of standard single-mode optical fiber and are analyzed by the Franson interferometer's arm 1. With our SPDC biphoton generation at ≈ 1316 nm, close to the zero-dispersion wavelength for standard single-mode optical fibers, there is little temporal walk-off caused by fiber dispersion. Thus, a dispersion-compensating module, which is normally lossy, is not required for the distribution of our high-dimensional BFC states. This entanglement distribution scheme is specially designed to be asymmetric according to the nature of the singly-filtered BFC. In the generation configuration of our singly-filtered BFC, only the signal photons suffer from the insertion and filtering loss of the FFPC. Thus, the idler photons with higher photon flux are suitable for long distance transmission and sent through the 10 km optical fiber with 3.63 dB transmission loss. Note that our entanglement distribution system is built using commercially available off-the-shelf components, and can be readily implemented in many existing quantum key distribution (QKD) systems.

We first measure the high-dimensional energy-time entanglement of the distributed singly-filtered BFC states via Franson interferometry. High-visibility Franson interference recurrences are observed even *after* 10 km distribution, with our singly-filtered BFC states. Figure 4b shows the measured Franson-interference visibilities at different relative delays after 10 km distribution, supported by our theoretical model. The insets of Figure 4b show the Franson fringes for different time bins, yielding a high visibility of 98.81 ± 0.61% at central time-bin with Bell inequality violation. In particular, compared to the Franson-interference recurrences in Figure 2c, the averaged degradation of the Franson interference visibility after 10 km distribution is only 1.21%. The high visibilities of the Franson-interference recurrences verify the successful distribution of genuine high-dimensional energy-time entanglement of our singly-filtered BFC. Furthermore, to quantitively lower bound the dimension of the distributed high-dimensional energy-time entanglement, we estimate the time-bin Schmidt number using the Franson interference recurrences in Figure 4b. In Figure 4b, we calculate the time-bin Schmidt number from the eigenvalues at each time bin and obtained a post-distribution Schmidt number of 12.99, with only 0.92% degradation compared to Figure 2e. Therefore, we lower bound the Hilbert-space dimensionality of the distributed time-binned singly-filtered BFC state to be at least 168 (= 12.99 × 12.99) [50].

Next we measure the energy-time entanglement between symmetric frequency-bin pairs of the singly-filtered BFC after 10 km asymmetric distribution. The frequency-binned energy-time entanglement is especially useful for wavelength-division multiplexing an entanglement-based quantum communication network [51]. Narrowband tunable BPFs are used to select different frequency bins for the signal and idler photons, cascaded with the Franson interferometer for energy-time entanglement measurement. Franson interference fringes for five symmetric frequency-bin pairs are observed as shown in the inset of Figure 4a. A high Franson interference visibility of 98.85 ± 0.50% after accidentals subtraction, is obtained for the central frequency-bin pair $S_0$&$I_0$ after 10 km asymmetric distribution. The Franson visibilities for frequency-bin pairs away from the central bin decrease as anticipated, but are still well above the quantum threshold to violate the Bell inequality. The frequency-binned Franson interference visibilities before and after 10 km asymmetric distribution are summarized in Figure 4c. An averaged Franson visibility of 96.70 ± 1.93% was obtained for all 5 frequency-bin pairs after distribution, with only 1.33% degradation compared to the near-distance frequency-binned energy-time entanglement verification described earlier (Figure 3). These high Franson interference visibilities certify the coherent transmission of the frequency-binned energy-time entanglement of our singly-filtered BFC. Simultaneously, we examine the coincidence counts of the central correlation peak (example noted in Figure 3c) for constructive interference, to extract the JSI of the singly-filtered BFC after 10 km asymmetric distribution. As summarized in Figure 4c, only the diagonal elements of the JSI have high coincidence counts, revealing the preserved frequency-bin entanglement after link distribution.

With the high-dimensional dual-basis platform of our singly-filtered BFC, we demonstrated high-dimensional QKD utilizing frequency multiplexed time-bin encoding based on the distribution setup. Our QKD measurements, however, were made *without* the 10-km-long fiber connection owing to an equipment limitation in our coincidence-counting setup that precluded accommodating that fiber's propagation delay. For key generation, Alice and Bob record a 10-second-long time stream of photon detections for post-detection processing, where their time streams are synchronized and divided into time-bins for discretization with variable bin durations. We use the layered low-density parity-check code to calculate the Shannon information upper bound on the photon information efficiency (PIE), the bits we can send per photon under highly-erroneous channel conditions [51]. The key rate is then obtained by multiplying each PIE with the number of photon pairs per second in the corresponding keystream. The total PIE of up to ≈ 14 bits/coincidence and total key rate of ≈ 4.7 kbits/s are obtained for 5 correlated frequency-bin pairs. We also compare the performance with the doubly-filtered BFC by relocating the FFPC before the PBS in Figure 4a and implementing the same QKD protocol. We observe ≈ 2.8× improvement in PIE and ≈ 49× improvement in raw key rate in Figure 4d, which benefit from the less filtering loss of our singly-filtered configuration. Secured by our Franson interference visibilities, we compute the upper bound on Eve's Holevo information from which the secure key rate can be obtained for the frequency-bin pairs [52]. As shown in Figure 4e, three frequency-bin pairs of our singly-filtered BFC achieve positive secure PIE after subtracting the upper bound of Eve's Holevo information. Utilizing the same protocol, we also calculate the secure key rate for the doubly-filtered BFC [13], and compare the two configurations in Figure 4f. Our singly-filtered BFC achieved a total secure key rate of 1.1 kbits/s, which still presents ≈ 7.5× improvement under the same system condition (photon source, detection system, etc.). Compared to the state-of-the-art QKD using high-dimensional encoding with several Mbit/s secure key rate [52], our scheme can be further improved by exploiting denser frequency multiplexing, brighter photon source and more efficient detectors. This proof-of-principle demonstration clearly illustrates the utility of our singly-filtered BFC for frequency-multiplexed QKD.

**Discussion**

To the best of our knowledge, this is the first demonstration of high-dimensional entanglement distribution of a singly-filtered BFC in both time and frequency domains. Quantum information encoded in time- and frequency-bins of entangled photons is naturally suitable for transmission over long distances using optical fibers or free-space links [1]. In particular, energy-time entanglement is ideally suited for high-dimensional QKD, with security monitored by the measurement of Franson visibility [52]. Franson and conjugate Franson interferometry [53], the latter of which has been recently demonstrated [45], can support the unconditional security of high-dimensional QKD systems. The long-distance distribution of frequency bins has been demonstrated using a dispersion cancellation system [31], with possible applications in wavelength multiplexed quantum networks. Our singly-filtered BFC, due to its inherent time-frequency entangled nature, can achieve high-dimensional encoding in both time- and frequency-bins, and our entanglement distribution demonstration exemplifies the singly-filtered BFC's potential as a practical high-dimensional entanglement platform for constructing future quantum networks.

We note that our singly-filtered BFC can be readily exploited to generate high-dimensional hyperentanglement by combining the polarization degree-of-freedom (DoF). With our type-II configuration, we can generate the post-selected polarization entanglement by mixing the signal and idler photons on a 50:50 beam splitter [10, 13]. Another common approach to generate polarization entanglement is using a Sagnac interferometer [39], which can also be adapted to suit our singly-filtered BFC. With polarization DoF, the dimensionality of our singly-filtered BFC can be doubled with an extra qubit encoded in the BFC photon pairs' polarization. High-dimensional hyperentangled BFC states provide a useful resource for applications, such as high-capacity quantum communication [54] and superdense quantum teleportation [55].

Our singly-filtered BFC platform also supports higher dimensional state generation. The dimensionality of frequency bins is determined by the bandwidth of the SPDC and the FSR of the FFPC. For example, using a SPDC source with 2 THz bandwidth and an FFPC with 10 GHz FSR will leads to a 200 × 200 dimensional system, which corresponds to 15 frequency-bin qubits. On the other hand, the BFC's time bins are the Fourier duals of the state's frequency bins, thus the scaling of time-frequency dimensionality is complementary. The time-bin and frequency-bin number product satisfies $N_t \times N_f = \pi B_{\text{SPDC}}/\Delta\omega$, where $B_{\text{SPDC}}$ is the SPDC bandwidth and $2\Delta\omega$ is the cavity linewidth [13]. Thus, the ideal dimensionality of the time-frequency space is determined by the FFPC linewidth with a given SPDC source. For example, with 2 THz SPDC bandwidth, 100 GHz cavity FSR and 1 GHz cavity linewidth, the time-bin number $N_t$ will increase to 100 with time-bin interval of 10 ps while the frequency-bin number will become $N_f = 20$, which leads to 13 time-bin qubits. The scaling of the BFC state generation can also be realized by integrating the Fabry–Pérot cavity with the SPDC source, which demonstrates frequency-multiplexed photon pairs covering 1,400 modes with 3.5 GHz FSR of the cavity [38]; or using an integrated microring resonator [11, 12, 31], where the achievable dimensionality is also bounded by the phase-matching bandwidth of the spontaneous four-wave mixing and the FSR of the microring resonator. The on-chip realization of high-dimensional BFCs can assist the examination of large-scale integrated quantum information processing [56].

The singly-filtered BFC will be particularly useful and readily implemented in a quantum repeater-based quantum internet [57]. In our singly-filtered configuration for BFC generation, the FFPC is external to the SPDC source, unlike the OPO approach. In a quantum repeater scheme with two entanglement sources [58], by inserting a Fabry-Pérot cavity in the path from each entanglement source to the Bell-state analyzer, the photons on the other path of both sources will be entangled in high-dimensions due to the singly-filtered configuration. This scheme enables possible high-dimensional quantum repeaters for use in long-distance quantum networks. Our singly-filtered BFC could thus provide a practical and flexible approach to increase the dimensionality of the time- and frequency-bases at a node of the quantum internet.

**Acknowledgements**

The authors acknowledge discussions with Zhenda Xie, Changchen Chen, Abhinav Kumar Vinod, Wei Liu, Jiahui Huang, Patrick Hayden, Pengfei Fan, Mi Lei, Charles Ci Wen Lim, and Alexander Euk Jin Ling; discussions on the superconducting nanowire single-photon detectors with Vikas Anant; Justin Caram for the help on the heralded auto-correlation measurements with a Hydraharp. This study is supported by the Army Research Office Multidisciplinary University Research Initiative (W911NF-21-2-0214), National Science Foundation under award numbers 1741707 (EFRI ACQUIRE), 1919355, 1936375 (QII-TAQS), and 2137984 (QuIC-TAQS). Part of this research was performed at the Jet Propulsion Laboratory, California Institute of Technology, under contract with NASA.

**Author contributions**

X.C. conceived the project and designed the experiments. K.-C.C. and X.C. conducted the measurements and performed data analysis. J.H.S., K.-C.C., X.C. and M.C.S. contributed to the theory and numerical modeling. A.M., M.S., M.D.S. and B.K. contributed the low-jitter SNSPD detectors. F.N.C.W, J.H.S., A.F. and C.W.W. supported and discussed the studies. X.C., K.-C.C., J.H.S. and C.W.W. prepared the manuscript. All authors contributed to the discussion and revision of the manuscript.

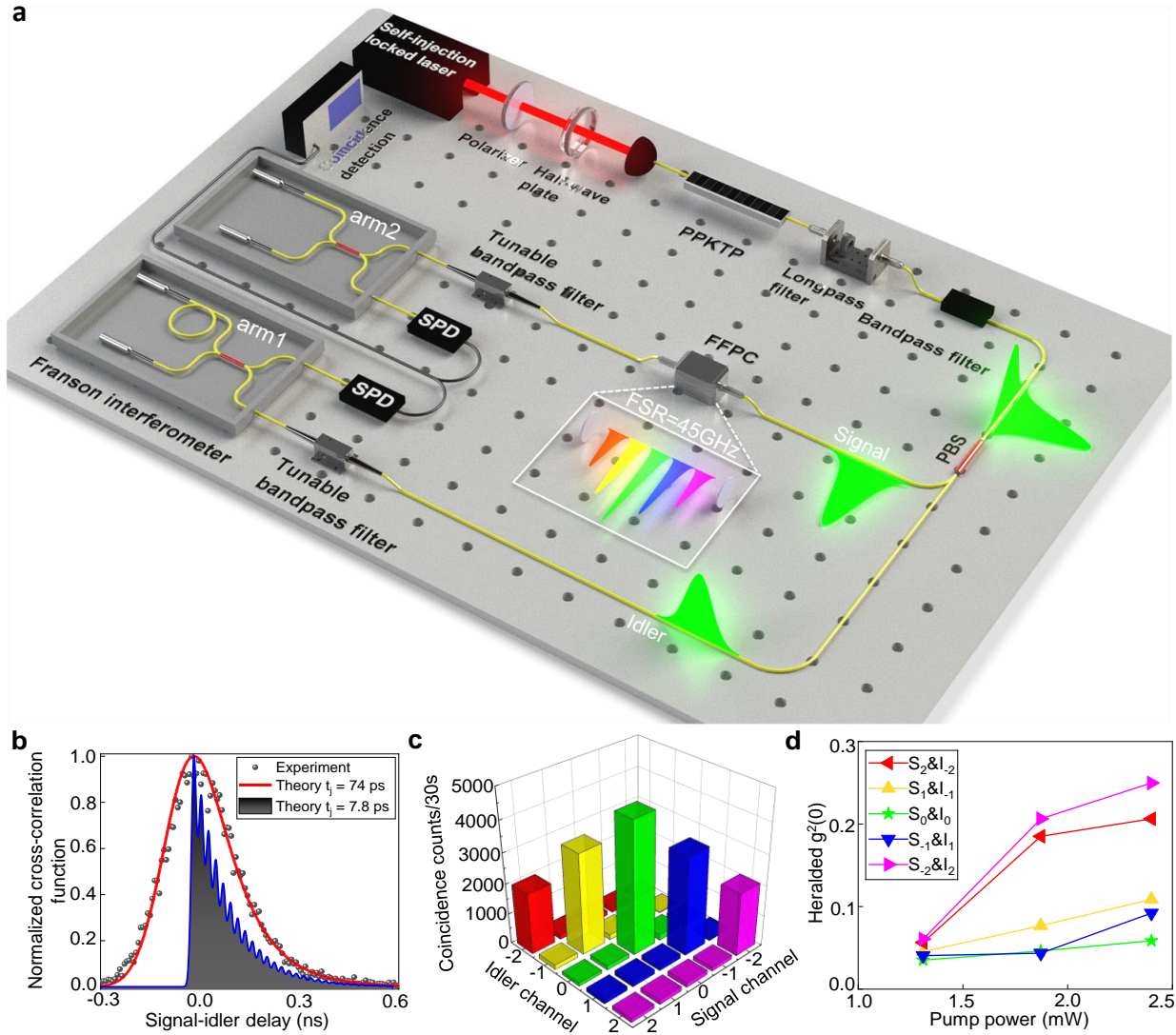

**Figure 1 | Generation and spectro-temporal characterization of singly-filtered biphoton frequency combs. a,** Experimental schematic for generation and characterization of a singly-filtered BFC. ppKTP: periodically-poled KTiOPO$_4$ waveguide; PBS, polarizing beam splitter; FFPC: fiber Fabry–Pérot cavity; SPD: superconducting nanowire single-photon detector. **b,** Experimental and theoretical temporal second-order cross-correlation function between signal and idler photons. Based on the 1.56 GHz cavity bandwidth and 45.32 GHz FSR of our FFPC, we theoretically fit our experimental results with different effective detector timing jitters. The periodic temporal oscillations of cross-correlation function for the 45.32 GHz cavity's singly-filtered BFC can be resolved when the effective timing jitter is set equal to or less than the cavity's round-trip time. **c,** Measured frequency-correlation matrix of our 45.32 GHz singly-filtered BFC within the 245 GHz SPDC bandwidth. The JSI is measured by using a pair of tunable bandpass filters with 300 pm bandwidths to select frequency bins for the signal and idler photons. Only the diagonal elements of the frequency-correlation matrix show high coincidence counts, revealing

the frequency-bin entanglement. **d,** Measured heralded single-photon second-order auto-correlation function $g^{(2)}(0)$ versus the pump power for each frequency-bin pair. The minimum heralded $g^{(2)}(0) \approx 0.035$ is measured for $S_0$&$I_0$ pair at 1.3 mW pump power with a heralding rate of 16 coincidences/s. All measured heralded $g^{(2)}(0)$'s are below the classical threshold, verifying the high purity of the frequency-binned single photon states from our singly-filtered BFC.

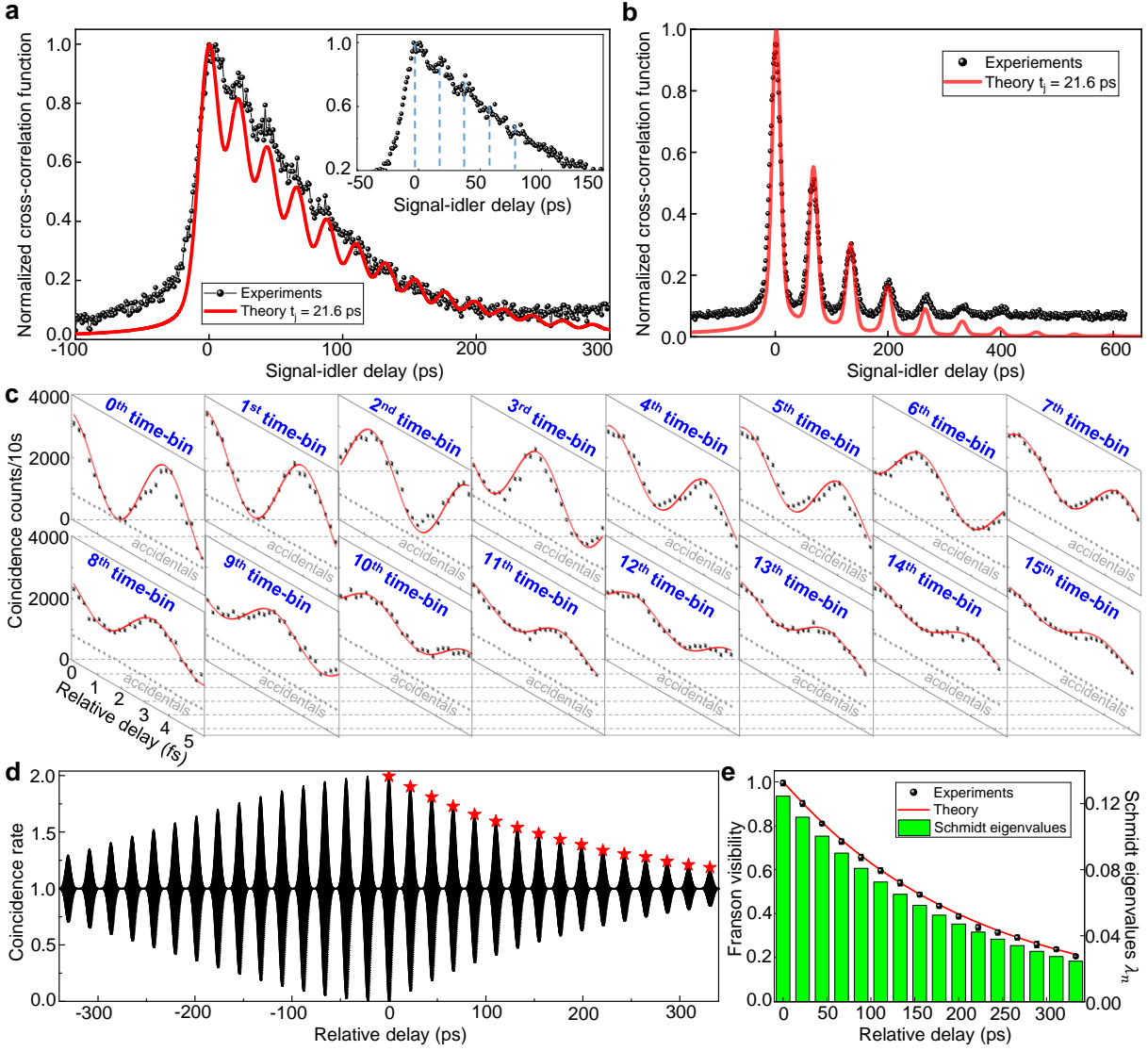

**Figure 2 | Temporal cross-correlations and non-local Franson examinations of high-dimensional energy-time entanglement of the singly-filtered BFCs. a,** Measured temporal second-order cross-correlation function between signal and idler photons and theoretical fitting with ≈ 21.6 ps combined timing jitter of the two differential SNSPDs – the timing jitter adds in quadrature. Temporal oscillation is resolved with a temporal period of 22.0 ps, which implies a cavity FSR of 45.45 GHz, matching well with our 45.32 GHz cavity. Inset: zoom in of the temporal second-order cross-correlation function with temporal spacing denoted by dashed lines. **b,** Measured second-order cross-correlation function with a 15.15 GHz cavity and theoretical fitting with ≈ 21.6 ps combined timing jitter of the two differential SNSPDs. The temporal spacing of the correlation peaks is fitted to be 66.8 ps, which corresponds to a cavity FSR of 14.97 GHz. **c,** Measured Franson-interference recurrences within the 360 ps traveling range of arm 1's delay line. The interference fringes are obtained from 0 (0$^{th}$ time-bin) to 15 (15$^{th}$ time-bin) round-trip times

of the 45.32 GHz FFPC, with the measured period of the interference recurrences found to be 22.1 ps. The fringe visibilities with (without) accidental coincidence counts subtracted from $0^{th}$ to $15^{th}$ time-bin are: 99.46 (72.34) ± 1.09% , 90.17 (64.13) ± 1.22%, 81.03 (51.05) ± 0.71%, 72.82 (47.06) ± 1.08%, 65.60 (43.23) ± 1.29%, 59.57 (40.37) ± 1.14%, 54.03 (38.02) ± 1.29%, 48.68 (34.39) ± 0.91%, 43.54 (30.22) ± 0.97%, 38.79 (27.08) ± 1.03%, 33.71 (25.44) ± 1.17%, 31.47 (24.41) ± 1.05%, 29.12 (19.79) ± 1.04%, 25.93 (18.82) ± 1.33%, 23.79 (16.85) ± 0.81%, 20.58 (15.07) ± 0.76%. The coincidence window for all the measurements is 2 ns. Error bars measured for each data point arise from Poisson statistics, experimental drift and measurement noise. **d,** Theoretical fringe envelope of Franson interference for the 45.32 GHz high-dimensional singly-filtered BFC, with superimposed experimental visibility results (red stars). **e,** Witnessed visibility of high-dimensional Franson interference fringes and Schmidt eigenvalues as a function of relative delay between arm 1 and arm 2. The theoretical visibilities, in percent, for the $n^{th}$ time-bin are 100, 90.44, 81.48, 73.57, 66.44, 59.93, 54.20, 48.93, 44.18, 39.87, 35.99, 32.49, 29.32, 26.47, 23.88 and 21.57, respectively. The Franson visibility decreases due to the FFPC's finite linewidth, as captured by our theory. The Schmidt eigenvalues are extracted from Franson-interference recurrences for 16 time bins, resulting in a Schmidt number lower bound of 13.11.

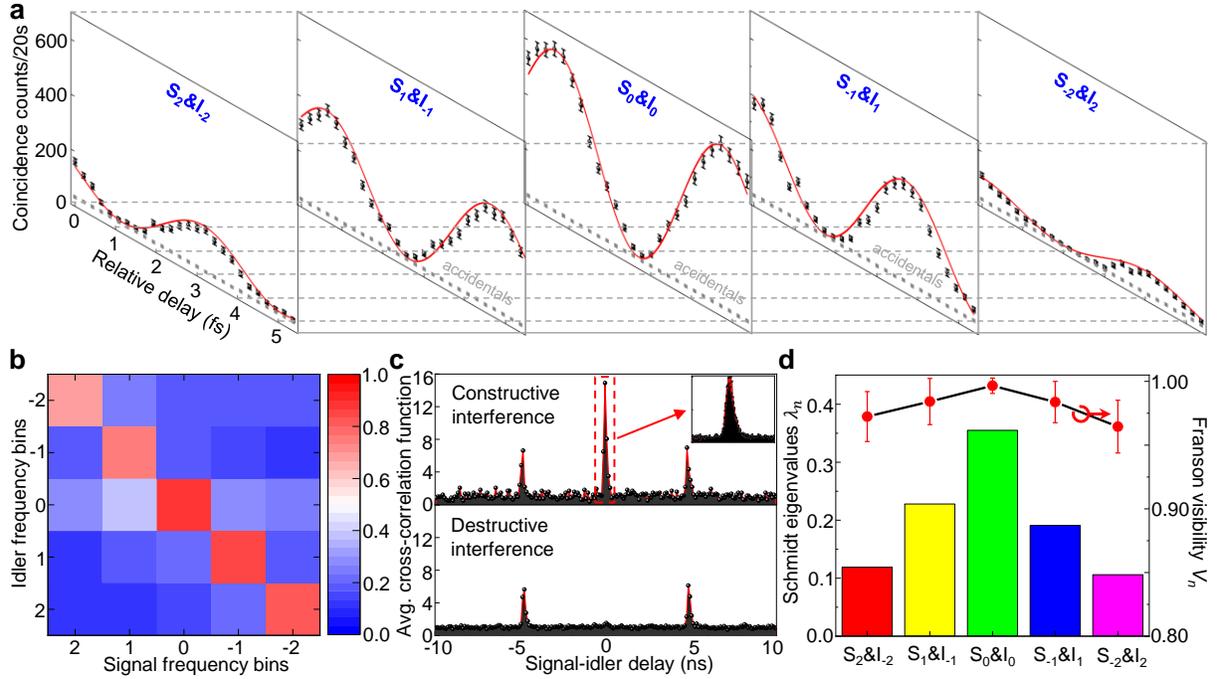

**Figure 3 | Frequency-binned energy-time entanglement verification via Franson interference.**
**a,** The measured Franson interference fringes, after accidentals were subtracted, for symmetric frequency-bin pairs. All the fringes were obtained at zero relative delay between arm 1 and arm 2 for optimum Franson interference. Maximum frequency-binned Franson interference is observed for $S_0$&$I_0$ pair with visibility up to $99.66 \pm 1.67\%$. The averaged frequency-binned Franson visibility for 5 pairs is $98.03 \pm 1.10\%$. The coincidence window for all the measurements was 2 ns. **b,** The Franson interference visibility map for frequency-bin pairs within the SPDC bandwidth. Frequency bins were selected using a pair of tunable BPFs with 300 pm bandwidths, that were manually tuned to scan from the -2 to +2 frequency bins from frequency degeneracy. Only the symmetric frequency-bin pairs in anti-diagonal terms show high Franson interference visibility. **c,** Recorded signal-idler cross-correlation function for constructive (top) and destructive (bottom) Franson interference. The inset is the zoom-in of the central correlation peak, showing the single-sided decay temporal signature of a singly-filtered BFC. **d,** Extracted Schmidt eigenvalues and Franson visibilities for 5 symmetric frequency-bin pairs. Frequency-binned Franson interference visibilities with (without) background subtracted are $97.24\ (82.88) \pm 1.96\%$ for $S_2$&$I_{-2}$, $98.42\ (90.90) \pm 1.81\%$ for $S_1$&$I_{-1}$, $99.66\ (92.06) \pm 0.60\%$ for $S_0$&$I_0$, $98.38\ (88.86) \pm 1.62\%$ for $S_{-1}$&$I_1$ and $96.45\ (81.53) \pm 2.07\%$ for $S_{-2}$&$I_2$, respectively. We extract the lower bound of the Schmidt number to be 4.17 across 5 symmetric frequency-bin pairs.

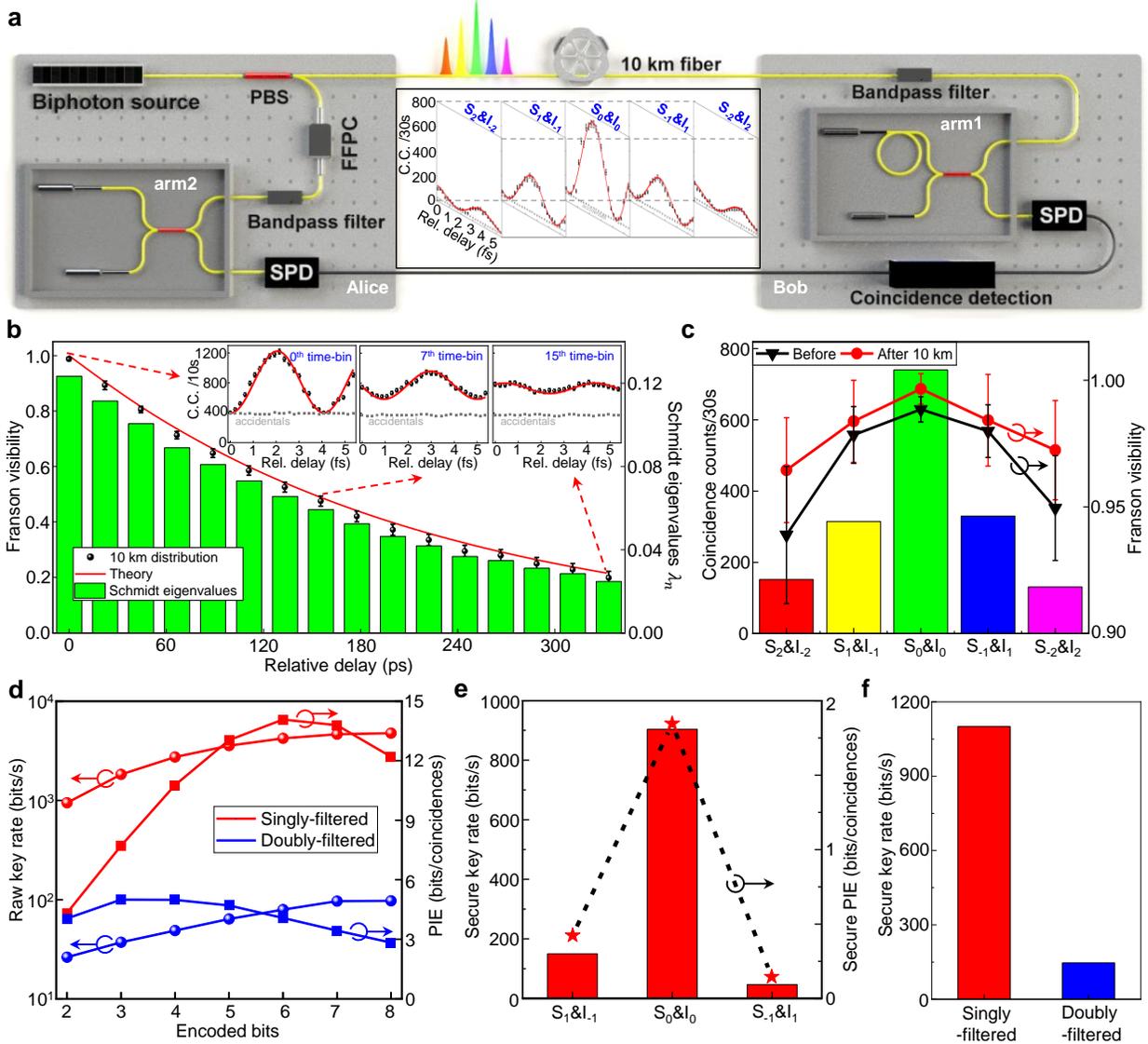

**Figure 4 | High-dimensional time-frequency entanglement distribution of a 45.32 GHz singly-filtered biphoton frequency comb at 10 km distances. a,** Illustrative experimental scheme for asymmetric entanglement distribution of high-dimensional time-frequency entanglement of a singly-filtered BFC. The singly-filtered BFC is generated by only passing the signal photons through an FFPC, while the idler photons are distributed via a 10 km fiber link. After distribution, a pair of tunable bandpass filters are used to select frequency bins for the signal and idler photons, which are then analyzed by a Franson interferometer for energy-time entanglement. The inset shows the Franson interference fringes for 5 symmetric frequency-bin pairs after 10 km distribution. **b,** The measured visibilities of Franson-interference recurrences after 10 km distribution from $0^{th}$ to $15^{th}$ time-bin with (without) background subtracted: 98.81 (63.90) ± 0.61%, 89.28 (55.29) ± 1.55%, 80.54 (49.44) ± 1.15%, 71.28 (40.10) ± 1.35%, 64.81 (32.90) ± 1.57%, 58.54 (28.83) ± 1.68%, 52.57 (20.92) ± 1.80%, 47.52 (19.87) ± 1.84%, 42.03 (17.40) ± 1.95%,

37.18 (14.67) ± 2.03%, 33.50 (14.62) ± 2.06%, 29.46 (13.78) ± 2.12%, 27.90 (13.64) ± 2.15%, 24.93 (13.06) ± 2.26%, 22.79 (11.21) ± 2.24%, 19.83 (10.79) ± 2.35%. The insets show the interference fringe for the $0^{th}$, $7^{th}$ and $15^{th}$ time-bin. The Franson interference visibility decay matches well with our theoretical prediction after 10 km distribution. The time-bin Schmidt number after 10 km distribution is measured to be 12.99, which only shows degradation of 0.92% compared to Figure 2e. **c,** The measured frequency-binned Franson-interference visibilities and the joint-spectral intensity measurements for 5 symmetric frequency-bin pairs of the singly-filtered BFC after 10 km distribution. Frequency-binned Franson interference visibilities after distribution with (without) background subtraction are 94.95 (73.86) ± 2.07% for $S_2$ & $I_{-2}$, 97.99 (84.12) ± 1.04% for $S_1$ & $I_{-1}$, 98.85 (87.60) ± 0.50% for $S_0$ & $I_0$, 97.84 (82.77) ± 1.12% for $S_{-1}$ & $I_1$ and 93.89 (70.76) ± 2.71% for $S_{-2}$ & $I_2$, respectively. Averaged frequency-binned Franson visibility of 96.70 ± 1.93% is obtained, with only 1.33% degradation after distribution. These results demonstrate that the frequency-binned energy-time entanglement of our singly-filtered BFC is well preserved after 10 km distribution. **d,** Raw key rate and photon information efficiency of the frequency-multiplexed quantum key distribution using the singly-filtered and doubly-filtered BFCs. **e,** Secure key rate and secure PIE at different frequency-bin pairs. Three frequency-bin pairs show positive PIE after subtracting Eve's Holevo information. A total SKR of 1.1 kbits/s is achieved using our singly-filtered BFC. **f,** Secure key rate comparison between singly-filtered BFC and doubly-filtered BFC. The singly-filtered BFC shows ≈ 7.5× improvement of secure key rate under the same system condition.